\documentclass{article}

\usepackage{arxiv}

\usepackage[utf8]{inputenc} 
\usepackage[T1]{fontenc}    
\usepackage{hyperref}       
\usepackage{url}            
\usepackage{booktabs}       
\usepackage{amsfonts}       
\usepackage{nicefrac}       
\usepackage{microtype}      
\usepackage{lipsum}

\usepackage{amsmath}
\usepackage{xcolor}
\usepackage{algorithm,algorithmic}
\usepackage{bm}
\usepackage{graphicx}

\title{Gaussian Process and Design of Experiments for Surrogate Modeling of Optical Properties of Fractal Aggregates}

\author{
  Ozan Burak Ericok \\
  Department of Mechanical Engineering\\
  Bogazici University\\
  Bebek, 34342, Istanbul, Turkey\\
   \And
 Atay Kaan Ozbek \\
 Department of Mechanical Engineering\\
  Bogazici University\\
  Bebek, 34342, Istanbul, Turkey\\
  \And
  Ali Taylan Cemgil \\
  Department of Computer Engineering\\
  Bogazici University\\
  Bebek, 34342, Istanbul, Turkey\\
  \And
  Hakan Erturk \thanks{Corresponding author\newline
  E-mail: hakan.erturk@boun.edu.tr (H. Erturk)}\\
  Department of Mechanical Engineering\\
  Bogazici University\\
  Bebek, 34342, Istanbul, Turkey\\
}

\begin{document}
\maketitle

\begin{abstract}
	A systematic approach based on the principles of supervised learning and design of experiments concepts is introduced to build a surrogate model for estimating the optical properties of fractal aggregates. The surrogate model is built on Gaussian process (GP) regression, and the input points for the GP regression are sampled with an adaptive sequential design algorithm. The covariance functions used are the squared exponential covariance function and the Matern covariance function both with Automatic Relevance Determination (ARD). The optical property considered is extinction efficiency of soot aggregates. The strengths and weaknesses of the proposed methodology are first tested with RDG-FA. Then, surrogate models are developed for the sampled points, for which the extinction efficiency is calculated by DDA. Four different uniformly gridded databases are also constructed for comparison. It is observed that the estimations based on the surrogate model designed with Matern covariance functions is superior to the estimations based on databases in terms of the accuracy of the estimations and the total number of input points they require. Finally, a preliminary surrogate model for $S_{11}$ is built to correct RDG-FA predictions with the aim of combining the speed of RDG-FA with the accuracy of DDA.

\end{abstract}


\maketitle

\section{Introduction}
	There are different analytical and computational methods to calculate the optical properties of nanoparticle systems. One of the most widely used methods to calculate the optical properties of irregularly shaped nanoparticles is the Rayleigh-Debye-Gans (RDG) approximation \cite{Kerker1969}. The Rayleigh-Debye-Gans approximation for fractal aggregates (RDG-FA) is formulated as an extension \cite{Sorensen2001,Faeth1995} that ignores the effects of both multiple and self-induced scattering, and it is valid if $\lvert m-1 \rvert  \ll 1 $ and $k \lvert m-1 \rvert d_p \ll 1$ where $m$, $k$ and $d_p$ are the refractive index, wavenumber and primary particle diameter of aggregate. Yon \textit{et. al} \cite{Yon2014} tried to include the multiple scattering effects into RDG-FA using rigorous calculations based on discrete dipole approximation (DDA) and generalized multi-sphere Mie-solution (GMM). They found that the size determination using static light scattering is not affected by multiple scattering effects. However, they suggested that the multiple scattering effects should be taken into account when absorption and extinction measurements are considered. Recently, Amin and Roberts \cite{Amin2017} used RDG-FA to retrieve some properties of soot aggregates such as volume fraction, primary particle diameter etc. from scattering and extinction measurements. They determined the primary particle diameter and the radius of gyration from scattering/extinction ratio and dissymmetry ratio, respectively. 
	
	The optical properties of irregularly shaped particles can also be calculated computationally. One of the fastest and  most accurate methods is the T-matrix method \cite{Waterman1965,Mackowski1996}, whose major limitation is that the surfaces of the spheres modeled must be non-overlapping. Another method that has no such limitation and frequently used is the discrete dipole approximation (DDA) introduced by Purcell and Pennypacker \cite{Purcell1973}. DDA models the particles by dipoles that are capable of interacting with both the incident field and each other. The dipole representation allows DDA to model very complex geometries with ease. The multiple and self-induced scattering effects are captured through the interactions of dipoles. Hence, DDA offers accurate estimations of the optical properties. More detailed information on DDA can be found in \cite{Draine1988}.
	
	The balance between accuracy and computational expense is the main concern for calculation of the optical properties of nanoparticle systems. While RDG-FA is one of the fastest approaches available, Ma \cite{Ma2011} compared angular scattering properties of soot aggregates comprised of 128 particles each having a diameter of 30 nm with $m=1.77+0.63i$, and observed a relative error up to 10\% in both $S_{11}$ and $S_{12}$ profiles calculated by RDG-FA and T-matrix methods. In another study, Zhao and Ma \cite{Zhao2009} studied the scattering cross section and some elements of Mueller matrix ($S_{11}$, $S_{12}$ and $S_{33}$) of soot aggregates, and observed a relative error up to 15\% in scattering cross section and up to 50\% in the Mueller matrix elements. They also stated that the applicability of RDG-FA exhibits a complicated pattern because of the refractive index dependence of the properties.  Talebi Moghaddam \textit{et. al} \cite{TalebiMoghaddam2017a} studied a similar problem and developed a methodology that combines the efficiency of RDG-FA with the accuracy of T-Matrix method by introducing an error function derived using principle component analysis. However, RDG-FA method is considered to be valid when $k \lvert m-1 \rvert d_p \ll 1$.  Beyond this range, methods such as DDA or T-matrix are considered to be very accurate when compared with RDG-FA but the computational time required to obtain optical properties is relatively high. For example, obtaining the orientation-averaged optical properties of a soot aggregate comprised of 16 nanoparticles each having 30 nm diameter at 532 nm wavelength by DDA using 3849 dipoles takes approximately 566 seconds in an 8 core system with 2.50 GHz frequency when 216 orientations are considered using DDSCAT, that is Fortran implementation of DDA \cite{Draine1994}.
	
	Therefore, some researchers preferred to create databases of optical properties of nanoparticle systems at discrete input points using higher accuracy models such as DDA, and using different interpolation methods to estimate the properties of a given nanoparticle system. For example, Menguc and Manikavasasagam \cite{Menguc1998} and Charnigo \textit{et. al} \cite{Charnigo2011} created databases to estimate elements of the Mueller matrix for soot aggregates and gold nanoparticles, respectively. Ericok \textit{et. al} \cite{Ericok2017,Ericok2018} created a monochromatic database for soot aggregates, and used cubic spline interpolations for estimations while solving the inverse characterization problem. Recently, they expanded their databases to multiple wavelengths to determine the characterization limits at different wavelengths \cite{Ericok2018a}. The main drawback of this approach is that the computational time required to create a fine gridded database might be significantly high, especially for higher dimensional problems. Considering that the optical properties of fractal aggregates depend on the direction, performing orientation averaging is also required for most of the problems of interest. Moreover, configuration averaging is also required while calculating the optical properties of fractal aggregates. Therefore, the optical properties of multiple aggregate realizations for multiple orientations should also be averaged while creating a database that significantly increases the computational effort required to built a database. 
	
	Surrogate models are fast mathematical or numerical representation of physical events based on sampled observations with known accuracy or uncertainty. They can replace accurate but computationally demanding physical models so that the response of the physical system can be estimated efficiently with a known uncertainty. Some of the widely used surrogate modeling techniques are polynomial surface response models \cite{Myers2009}, kriging (also known as Gaussian process) \cite{Cressie1988,Rasmussen2006}, radial basis functions \cite{Dyn1986}, support vector regression \cite{Clarke2005} and artificial neural networks \cite{Yegnanarayana2009}.  
	
	The accuracy of a surrogate model is directly related to the observations it relies on that are referred as training points. The training points are comprised of responses or outputs of a system for given a condition or input.  Hence, developing a surrogate model is determination of a representation so that given the input, the corresponding output points can be estimated with a desired accuracy. While it can be presumed that more observations would lead to more accurate surrogate models, the cost required to obtain an observation is also an important factor.  Considering the cost of experimental work, training points are often generated using accurate numerical simulations that take multiple parameters into account for an observation and can use considerable computation time. Therefore, efficient sampling strategies are generally used to build surrogate models with the least amount of observations possible and the sampling procedure that is also known as design of experiments is a critical stage of developing an accurate surrogate model. The most basic approach might be to sample uniformly distributed points from the input domain. While this is a very straightforward approach, it would require higher number of samples for accurate representation if the response of the physical system under investigation is not smooth, or have multiple extrema.  In such situations, it would be preferable to sample higher number of points from regions where the response has steep gradients so that all the variations in the output domain can be captured. This can be possible by sequential sampling rather than all at once that is known as one-shot sampling. The sequential sampling approaches can be classified as space-filling sequential sampling and adaptive sampling. In space-filling sequential sampling, the main objective is to sample points from the input domain as evenly as possible, and to cover the entire input domain. Since the points are sampled one at a time the methods based on this approach do not require a predetermined number of sampling points. The sequential space-filling approaches do not consider the outputs of the sampled points and might not fully capture the topology of the output domain. On the other hand, adaptive sequential sampling (also known as active learning) approaches consider the results from the simulations, and they adaptively learn the underlying function as more points are sampled. Thus, more points are sampled from the interesting regions and the input domain is covered more efficiently using minimum number of samples. The adaptive sampling approaches become more suitable when each simulation is costly as they potentially require fewer points than space-filling sequential sampling. More information on this subject and example use of these approaches can be found in excellent reviews of Liu \textit{et al.} \cite{Liu2018} and Garud \textit{et al.} \cite{Garud2017}. 
	
	In this study, the fundamental aim is to provide a proper methodology to build an accurate surrogate model based on Gaussian process regression for predicting the extinction efficiency of nanoparticle aggregates.  An adaptive sequential sampling strategy is adopted for the selection of the observations that are calculated by DDA so that the number of observations required can be minimized.  The strengths and weaknesses of the proposed method are investigated with respect to interpolating from a database and predictions by RDG-FA. 

\section{Theory}

\subsection{\label{sec:fractal_aggregates}Fractal Aggregates}
	
		Nanoparticle aggregates are generally represented with the fractal equation,
		
		\begin{equation}
		N_p = k_f { \Bigg( \frac {2R_g} { d_p } \Bigg) } ^ {D_f}
		\label{eq:eq_fractal}
		\end{equation}
		where $N_p$ is the number of particles in the aggregate, $d_p$ is the primary particle diameter, $R_g$ is the radius of gyration. $D_f$ and $k_f$ are the fractal dimension and fractal prefactor, respectively.
		
		In most cases, an aggregate does not have to be comprised of primary particles of the same size. Primary particle diameter of soot aggregates, for example, generally follows a lognormal distribution,
		
		\begin{equation}
		p(d_p) = \frac{1}{d_p \sqrt{2 \pi} \ln \sigma} \exp{\Bigg[- \Bigg( \frac{\ln {(d_p/d_{geo})}}{\sqrt{2} \ln \sigma} \Bigg)^2 \Bigg]}
		\label{eq:dp_lognormal}
		\end{equation}
		where $d_{geo}$ and $\sigma$ are geometric mean and geometric standard deviation of primary particles.
		
		Filippov's algorithm \cite{Filippov2000} is one of the most widely used methods to generate fractal-like aggregates in the literature. There are mainly two approaches to generate fractals. The first one is known as  particle cluster algorithm that adds particles one by one to a growing cluster while satisfying the fractal equation at every step. The second approach is known as cluster-cluster aggregation algorithm, where smaller aggregates are first formed, then combined to generate larger aggregates. Although the implementation of particle cluster algorithm is more straightforward, it was reported that it fails to produce the expected slope of pair-correlation function  \cite{Filippov2000,Skorupski2014}. Therefore, Filippov’s cluster-cluster aggregation algorithm is used in this study to generate non-overlapping nanoparticle aggregates. While the imposed fractal dimension is not conserved for an individual aggregate using this algorithm, it is conserved for the ensemble \cite{Moran2019}.  Hence, this algorithm is used with ensemble averaging to overcome this problem.

	\subsection{\label{sec:radiative_properties}Radiative Properties of Nanoparticle Aggregates}
	
		One of the most widely adopted methods to calculate the radiative properties of nanoparticle aggregates made up of same particle size is the RDG-FA due to its simplicity and speed. The RDG-FA approximates the total absorption and scattering cross sections of monodisperse aggregates as \cite{Sorensen2001}
		
		\begin{equation}
		C_{sca}^{agg} = N_p^2 C_{sca}^{mon} G(k R_g)
		\label{eq:rdg_sca_cross_section_mono}
		\end{equation}
		\begin{equation}
		C_{abs}^{agg} = N_p C_{abs}^{mon}
		\label{eq:rdg_abs_cross_section_mono}
		\end{equation}
		where $k$ is the wavenumber and 
		
		\begin{equation}
		G(k R_g)=\Bigg( 1 + \frac{4}{3 D_f} k^2 R_g^2 \Bigg) ^ {-D_f / 2}
		\label{eq:scattering_factor}
		\end{equation}		
		is the scattering factor. 
		
		\begin{equation}
		C_{sca}^{mon} = \frac{\pi}{24} k^4 d_p^6 F(m)
		\label{eq:scat_cross_sec_mon}
		\end{equation}
		and 
		
		\begin{equation}
		C_{abs}^{mon} = \frac{\pi}{2} k d_p^3 E(m)
		\label{eq:abs_cross_sec_mon}
		\end{equation}
		represents the scattering and absorption cross sections of an individual monomer, respectively. Here,
		
		\begin{equation}
		F(m) = \left| \frac{m^2-1}{m^2+2} \right| ^2
		\label{eq:f_function}
		\end{equation}
		and 
		
		\begin{equation}
		E(m)=\text{Im} \Big( \frac{m^2-1}{m^2+2} \Big)
		\label{eq:e_function}
		\end{equation} 
		are functions of complex refractive index of the material.
		
		Liu \textit{et. al} \cite{CLiu2015} proposed two different adjustments for the aggregates with particle size distribution, and found that the one using effective diameter, $d_{eff}$, is superior to the other. The scattering and absorption cross sections of aggregates with different-sized monomers are defined accordingly as
		
		\begin{equation}
		C_{sca}^{agg} = N_p^2 \frac{\pi}{24} k^4 d_{geo}^6 \left[ \exp(4.5 \ln^2 \sigma) \right]^2 F(m) G(k R_g, D_f)
		\label{eq:rdg_sca_cross_section_poly}
		\end{equation}
		
		\begin{equation}
		C_{abs}^{agg} = N_p \frac{\pi}{2} k d_{geo}^3 \exp(4.5 \ln^2 \sigma) E(m)
		\label{eq:rdg_abs_cross_section_poly}
		\end{equation}
		
		Once the scattering and absorption cross sections are calculated the extinction cross section, $C_{ext}^{agg}$, can be computed as
		
		\begin{equation}
		C_{ext}^{agg} = C_{sca}^{agg} + C_{abs}^{agg} 
		\label{eq:rdg_ext_cross_section_poly}
		\end{equation}
		
		The scattering, absorption and extinction efficiencies, $Q$, can be calculated by
		
		\begin{equation}
		Q_i^{agg} = C_i^{agg} \bigg/ {(\pi {d_{eff}^2}/{4})} \quad\quad\quad i=sca,abs,ext
		\label{eq:optical_efficiencies}
		\end{equation}
		where the effective diameter, $d_{eff}$, can be defined based on the total volume of the aggregate as 
		
		\begin{equation}		
		V=\pi d_{eff}^3 / 6
		\label{eq:eff_dia_1}
		\end{equation}
		and can be calculated using Eq. \ref{eq:dp_lognormal} as 
		
		\begin{equation}
		d_{eff} = d_{geo} \left[ N_p \exp(4.5 \ln^2 \sigma) \right] ^ {1/3}
		\label{eq:eff_dia_2}
		\end{equation}

        The scattering event is fully characterized by a $4\times4$ matrix known as the \textit{Mueller matrix}, $S$, \cite{Bohren1998}.
        
        \begin{equation}
        \begin{bmatrix} \mathbf{K}_s
        \end{bmatrix}
        =
        \frac{1}{k^2r^2}
         \begin{bmatrix}
            S
          \end{bmatrix}
       \begin{bmatrix}
        \mathbf{K}_i
      \end{bmatrix}
		\label{eq:Mueller}
		\end{equation}
	    where $r$ is the distance between the detector and the scatterer, and $\mathbf{K}_i$ and $\mathbf{K}_s$ are the incident and scattered intensity vectors, respectively.
		
		The functional dependence of Mueller matrix elements on the scattering angle $\theta$ allows us to calculate the directional properties of the scattering event. If the incident light is unpolarized, the differential cross section, ${dC_{sca}^{agg}}/{d\Omega}$, of an optical element can be calculated by \cite{Draine08}.
		
		\begin{equation}
	    \frac{dC_{sca}^{agg}}{d\Omega} = \frac{1}{k^2}S_{11}
		\label{eq:diff_sca_cross_section_Mueller2}
		\end{equation}
		where $S_{11}$ is the first element of the Mueller matrix.
		
	The differential scattering cross section of an aggregate can also be calculated using RDG-FA method with an appropriate structure factor. Then, Eq. \ref{eq:diff_sca_cross_section_Mueller2} can be used to estimate $S_{11}$ predicted by RDG-FA. In order to take into account the polydispersity in the primary particle diameter, the differential scattering cross section given by Sorensen \cite{Sorensen2001} is combined with the adjustment proposed by Liu \textit{et al.} \cite{CLiu2015}, and the following equation is derived.
	   
	    \begin{equation}
		\frac{dC_{sca}^{agg}}{d\Omega} = N_p^2 k^4 (d_{geo}/2)^6 \left[ \exp(4.5 \ln^2 \sigma) \right]^2 F(m) S(q R_g)
		\label{eq:diff_sca_cross_section_poly}
		\end{equation}
    where $S(qR_g)$ is the structure factor function, and $q$ is the magnitude of the scattering wave vector, $q=4\pi\lambda^{-1} sin(\theta/2)$. The structure factor for polydisperse aggregates considered in this study is adapted from \cite{Dobbins91}.

	    \begin{equation}
		S(q R_g)=\exp[-(qR_g)^2/3] \qquad \qquad qR_g<\sqrt{3D_f/2}
		\label{eq:st_fac_Dobbins_1}
		\end{equation}
		
	    \begin{equation}
		S(q R_g)=\Bigg(\frac{3D_f}{2e}\Bigg)^\frac{D_f}{2}(qR_g)^{-D_f} \qquad \qquad qR_g\geq\sqrt{3D_f/2}
		\label{eq:st_fac_Dobbins_2}
		\end{equation}
		where $e$ is the base of natural logarithm.

\section{Gaussian Process (GP)}

A physical model is a mathematical mapping between applied boundary conditions and the corresponding responses of a system. While building surrogates, the observations are used for training as the surrogate model ``learns'' with the observed data and mimics the actual physical model. This process is known as supervised learning, and it is called regression for continuous outputs and classification for discrete outputs. Once an accurate surrogate model is built, the outputs for different input points that are not included in the training set can be predicted easily. In this study, the surrogate model is built based on Gaussian process regression.
	
	Considering a general system described by multiple parameters, the parameter set defining the system can be represented as an input vector, $\mathbf{x}$, and the corresponding behaviour of the system can be denoted by a scalar output variable, $y$.  Suppose that a training set $\mathcal{D}$ consists of $N$ observations, $\mathcal{D} = (\mathbf{X},\mathbf{y}) = \{ (\mathbf{x}_i,y_i) | i=1:N \}$.  The fundamental aim of a surragate model is to predict $y^\ast$ for a new input $\mathbf{x}^\ast$ given the observations $\mathcal{D}$.  This aim necessitates to make some assumptions about the underlying function relating an input to the corresponding output. A general approach is to consider a set of orthogonal functions and try to fit the observations to the function set. This approach could be problematic when the function set considered is not capable of modeling the data well. Alternatively, one can assign prior probabilities to each function rather than selecting a specific function set, and the functions that represent the data better will get higher probabilities. However, assigning probabilities to every function could take a significant time. One possible solution to this problem is to use Gaussian process that can be used to infer functions.  
	
	Consider that the observations are related to the inputs with an unknown function, $f$, where $y_i=f(\mathbf{x}_i)$, and consider also that the observations are noisy. One can infer a distribution over functions based on the observations, and then use it to calculate the predictions for new inputs as,
	\begin{equation}
	p(y^\ast \vert \mathbf{x}^\ast, \mathbf{X}, \mathbf{y}) = \int p(y^\ast \vert f, \mathbf{x^\ast}) p(f \vert \mathbf{X}, \mathbf{y}) df
	\label{eq:gp_general}
	\end{equation}
	
	A Gaussian process (GP) is a generalization of Gaussian probability distribution for inferring functions, and it is a stochastic process defining probabilities over functions rather than scalars or vectors \cite{Rasmussen2006,Murphy2012}. A GP assumes that the probability of the observations, $p(y_1, ..., y_N)$, is jointly Gaussian with some mean, $m(\mathbf{x})=\mathbb{E}[ f(\mathbf{x}) ]$, and covariance, $k(\mathbf{x},\mathbf{x'})=\mathbb{E}[ (f(\mathbf{x}) - m(\mathbf{x}) ) (f(\mathbf{x'}) - m(\mathbf{x'}) )^T]$, functions where $\mathbb{E}$ is the expectation operator. Then, the GP is simply written as,
	
	\begin{equation}
	f(\mathbf{x}) \sim \mathcal{GP}\big(  m(\mathbf{x}),k(\mathbf{x},\mathbf{x'}) \big)
	\label{eq:gp_definition}
	\end{equation}

	\subsection{\label{sec:noisy_predictions}Predictions using noisy observations}
		Consider that an observed measurement set $\mathcal{D} = (\mathbf{X},\mathbf{y}) = \{ (\mathbf{x}_i,y_i) | i=1:N \}$ with $\mathbf{x} \in \mathcal{R}^D$ is noisy and modeled as $y=f(\mathbf{x}) + \epsilon$ where $\epsilon$ is the additive error that is generally modeled as Gaussian noise with a variance of $\sigma_n^2$, $\epsilon \sim \mathcal{N}(0,\sigma_n^2)$. In such a case, the covariance of the observed noisy data is,
		\begin{equation}
		\text{cov}(y_p, y_q) = k_y(\mathbf{x}_p, \mathbf{x}_q) = k(\mathbf{x}_p, \mathbf{x}_q) +  \sigma_n^2 \delta_{pq} 
		\label{eq:gp_covariance_function_definition}
		\end{equation}
		where $\delta_{pq}$ is the Kronecker delta function. Equation \ref{eq:gp_covariance_function_definition} can also be represented as,
		\begin{equation}
		\text{cov}(\mathbf{y}) = \mathbf{K}_y = \mathbf{K} +  \sigma_n^2 \mathbf{I}
		\label{eq:gp_covariance_kernel_definition}
		\end{equation}
		where $\mathbf{K}=k(\mathbf{X},\mathbf{X})$ and $\mathbf{I}$ are the positive-definite covariance and identity matrices of size $N \times N$, respectively. 
		
		The aim is to predict the noise-free outputs, $\mathbf{f}^\ast$, for a given test set, $\mathbf{X}^\ast$, of size $N^\ast \times D$ using the noisy observation set $\mathcal{D}$. The joint distribution of the training outputs, $\mathbf{y}$, and the noise-free test outputs, $\mathbf{f}^\ast$, can be written as,
		\begin{equation}
		\begin{bmatrix}
		\mathbf{y} \\ \mathbf{f^\ast}
		\end{bmatrix}
		\sim
		\mathcal{N}\left(\mathbf{0}, 	
		\begin{bmatrix}
		\mathbf{K}_y		&  \mathbf{K}^\ast\\ 
		\mathbf{K}^{\ast T}	&  \mathbf{K}^{\ast\ast}\\ 
		\end{bmatrix} \right)
		\label{eq:gp}
		\end{equation}
		where $\mathbf{K}^\ast = k(\mathbf{X},\mathbf{X}^\ast)$ and $\mathbf{K}^{\ast\ast} = k(\mathbf{X}^\ast,\mathbf{X}^\ast)$ of sizes $N \times N^\ast$ and $N^\ast \times N^\ast$, respectively. It should be noted that zero mean is assumed for notational simplicity. Derivation of the posterior distribution of $\mathbf{f}^\ast$ yields the predictive equations of GP regression as,
		\begin{align}
		&p(\mathbf{f}^\ast | \mathbf{X}, \mathbf{y}, \mathbf{X}^\ast)  =  \mathcal{N} \big( \mathbf{f}^\ast | \bm{\mu}^\ast, \mathbf{\Sigma}^\ast    \big)\\
		&\bm{\mu}^\ast = \mathbf{K}^{\ast T} \mathbf{K}_y^{-1} \mathbf{y} \\
		&\mathbf{\Sigma}^\ast = \mathbf{K}^{\ast\ast} - \mathbf{K}^{\ast T} \mathbf{K}_y^{-1} \mathbf{K}^\ast
		\label{eq:gp_predictive}
		\end{align}

	\subsection{\label{sec:covariance_functions_hyperparameters}Covariance Functions and Hyperparameters}
		According to GP; outcomes of two points $\mathbf{x}_p, \mathbf{x}_q \in \mathcal{R}^D$ are similar if the covariance function deems these points to be similar. Therefore, the choice of a covariance function is critical. A typical example for the covariance function $k(\mathbf{x}_p, \mathbf{x}_q)$ defined in Eq. \ref{eq:gp_covariance_function_definition} can be the squared exponential covariance function with automatic relevance determination (ARD) distance measure that is defined as
		\begin{equation}
		k(\mathbf{x}_p, \mathbf{x}_q) = \sigma_f^2 \, \exp \big( - \frac{ (\mathbf{x}_p - \mathbf{x}_q)^T \mathbf{\Lambda}^{-2}  (\mathbf{x}_p- \mathbf{x}_q)}{2}     \big)
		\label{eq:gp_covSEard}
		\end{equation}
		where $\sigma_f^2$ is the signal variance. $\mathbf{\Lambda}$ is a diagonal matrix with ARD parameters $\lambda_1,\lambda_2 ... \lambda_D$. The independent variables $(\sigma_f, \lambda_1,\lambda_2 ... \lambda_D)$ that need to be identified to define the covariance functions are known as \textit{hyperparameters}, and they have significant impact on the quality of the prediction.
		
		The optimal values of hyperparameters are generally estimated by maximizing the marginal likelihood that is defined as,
		\begin{equation}
		p(\mathbf{y} | \mathbf{X}) = \int p( \mathbf{y} | \mathbf{f}, \mathbf{X} ) p( \mathbf{f} | \mathbf{X} ) d\mathbf{f}
		\label{eq:gp_marginal_likelihood}
		\end{equation}
		
		Finally, the log marginal likelihood can be calculated as,
		\begin{equation}
		\text{log} \ p( \mathbf{y} | \mathbf{X} ) = - \frac{1}{2} \mathbf{y}^T \mathbf{K}_y^{-1} \mathbf{y} - \frac{1}{2} \text{log} |\mathbf{K}_y| - \frac{n}{2} \text{log}(2 \pi)
		\label{eq:gp_log_marginal_likelihood}
		\end{equation}

\section{Problem Statement}

Many applications involving radiative transfer equation require optical cross sections or efficiencies of nanoparticle systems. Talebi Moghaddam \textit{et. al} \cite{TalebiMoghaddam2016} studied the changes in local absorption efficiencies of gold nanoparticle systems when an atomic force microscope tip is placed very close to the particles. Bond and Bergstrom \cite{Bond2006a} studied the absorption of soot aerosol systems to investigate the effects of these systems on climate. Hadwin \textit{et. al} \cite{Hadwin2018} used RDG-FA to calculate the absorption efficiencies of soot aggregates in time-resolved laser-induced incandescence experiments. Kandilian \textit{et. al} \cite{Kandilian2015} studied absorption, scattering and extinction cross sections of microalgae aggregates and proposed a coated sphere approximation to calculate these properties. Xu \textit{et. al} \cite{Xu2013} calculated extinction cross section and the differential scattering cross section of polydisperse gold nanorod ensembles with T-matrix method, and used these properties to find the critical geometric parameters of gold nanorods. Recently, Tazaki \textit{et. al} \cite{Tazaki2018} compared the absorption, scattering and extinction cross sections of agglomerates generated with ballistic cluster cluster and ballistic particle cluster aggregation using RDG, T-matrix, effective medium theory and the distribution of hollow sphere method. Following these studies, this study focuses on modeling the extinction efficiency of fractal soot aggregates.
	
	The measurement ensembles considered in this work are assumed to be comprised of aggregates with same number of nanoparticles having a lognormal particle size distribution. Previous experimental studies \cite{Koylu1992,Bescond2014,Cenker2015a} showed that the primary particle radii of soot aggregates follows a lognormal distribution with primary particle diameter varying between 5-60 nm and geometric standard deviation varying up to 1.54. In this study, the primary particle diameter and the geometric standard deviation are constrained with $d_p \in \left[  6 \,\, 54 \right]$ nm and $\sigma \in \left[  1.0 \,\,  1.5 \right]$. Tian \textit{et. al} \cite{Tian2004} experimentally showed that the highest probability density region of $N_p$ covers 5-60. Therefore, it is also assumed that $N_p \in \left[  6 \,\, 54 \right]$ for the training set in this study. The remaining region in considered for test points. Moreover, the fractal aggregates are generated using Filippov's cluster-cluster algorithm \cite{Filippov2000,Skorupski2014}. The optical properties are averaged over 216 directions. The fractal parameters $D_f$ and $k_f$ are taken as 1.81 and 1.37, respectively. The optical properties of soot are calculated at $\lambda=532$ nm. Refractive index of soot aggregates at 532 nm is taken from the study of Yon \textit{et. al} as $m=1.61+i0.74$ \cite{Yon2011}. 
	
	Finally, the training dataset $\mathcal{D} = \{ (\mathbf{x}_i,y_i) | i=1:N \}$ contains $N$ input points $\mathbf{x} \in \mathcal{R}^3$ with $\mathbf{x} = \left[ N_p \, d_p \, \sigma \right]$. The outputs $y$ are the extinction efficiencies of soot aggregates calculated with either RDG-FA or DDSCAT.  T-matrix method could have been used in place of DDSCAT to increase computational efficiency, considering that the particle overlapping, and necking effects are ignored in this study. However, DDSCAT is preferred as as this method enables consideration of these effects in further studies.

\section{Methodology}
\begin{figure}[h!]
		\centering
		\fbox{\includegraphics[width=\linewidth]{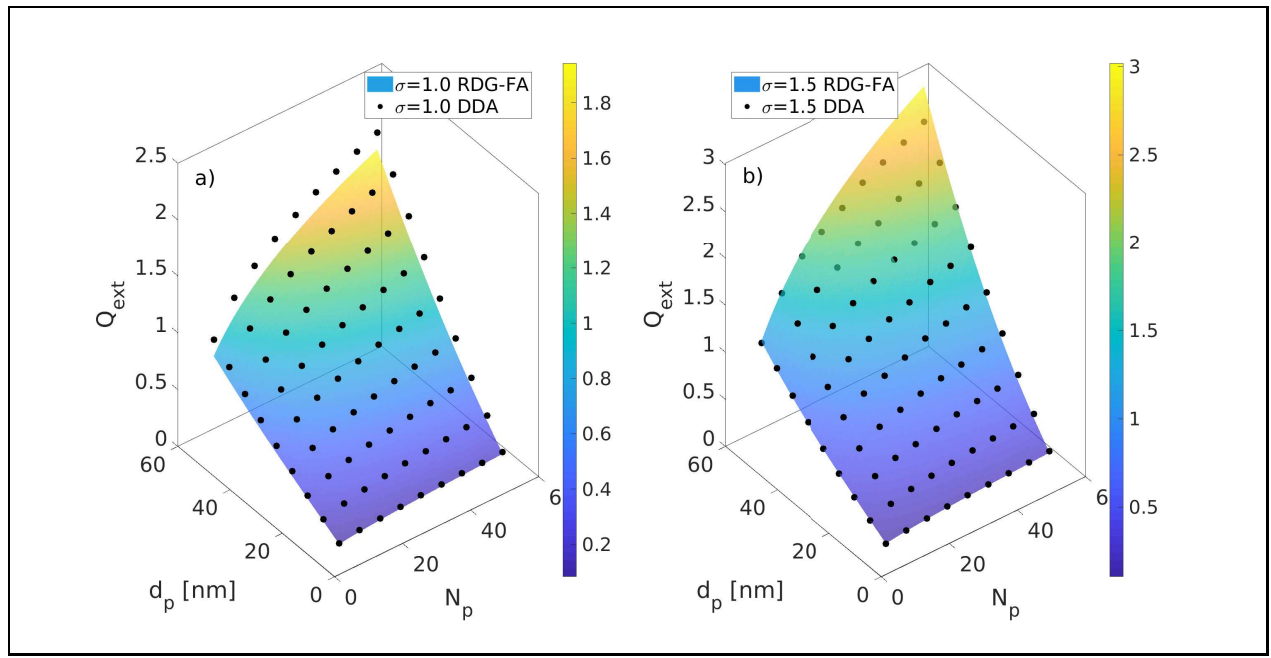}}
		\caption{$Q_{ext}$ calculated by RDG-FA theory and DDA as a function of $N_p$ and $d_p$ for a) $\sigma=1.0$ and b) $\sigma=1.5$. The surfaces show RDG-FA predictions wheres the dots show DDSCAT predictions.}
		\label{fig:fig1}
	\end{figure}

The objective is to develop an accurate surrogate model of extinction efficiencies for fractals based on GP that is satisfying a desired tolerance level, minimizing the number of training points calculated by DDSCAT. Therefore, an adaptive sequential design algorithm is used to sample training points for the surrogate model. Moreover, the surrogate models are built with different covariance functions to identify their capability.  A three step procedure is followed for building the surrogate models  for $Q_{ext}$. In the first step, the adaptive sequential design algorithm is utilized for identifying sampling points using RDG-FA as it provides fast approximations of the optical properties. Multiple simulations are conducted to determine the optimum parameters of the adaptive sequential design algorithm.  In the second step, surrogate models with different covariance functions are built for the input points sampled in the first step.  The extinction efficiencies of the input points in this step are also predicted by RDG-FA, due to its efficiency. Figure \ref{fig:fig1} shows $Q_{ext}$ surfaces calculated with RDG-FA and DDSCAT as a function of $N_p$ and $d_p$ for two different geometric standard deviation values, $\sigma=1.0$ and $\sigma=1.5$.  It can be observed that although RDG-FA produces relatively erroneous estimates, it captures the overall trend reasonably, and therefore, can be used for determination of the input points and covariance functions. In the third step, the input points identified using RDG-FA are re-evaluated by DDSCAT and these samples are used to built a surrogate model.
	
	Once the surrogate model for $Q_{ext}$ is built, a surrogate model for $S_{11}$ is developed next.  For that, the $S_{11}$ values that are predicted along with the $Q_{ext}$ values using RDG-FA and DDSCAT for the sample training points mentioned above are used. The error in the RDG-FA predictions are then calculated and used to train a surrogate model for $S_{11}$ in conjuction with the predictions of RDG-FA. The calculations regarding the surrogate model based on GP are conducted using the Gaussian Process for Machine Learning (GPML) Toolbox developed by Rasmussen and Nickish \cite{Rasmussen2010}.  
	
	\subsection{\label{sec:adaptive_sequential_design_algorithm}Adaptive Sequential Design Algorithm}
Adaptive sequential design algorithms intend to reduce the number of sample points required with their exploration and exploitation strategies. These two strategies are complementary to each other; while the former tends to sample points as evenly as possible to cover the input domain, the latter tends to identify more topologically interesting regions of the output domain and samples more points around those regions. An efficient adaptive sequential design algorithm should find a balance between its exploration and exploitation strategies. In this work, the adaptive sequential design algorithm (Algorithm \ref{algorithm:adaptive_sequential_design_algorithm}) proposed by Ajdari and Mahlooji \cite{Ajdari2014} is adopted, and it is briefly explained in this section.
		
		The algorithm starts with the initial sampling of points that are generated by Latin Hypercube Design (LHD), based on the number of initial points, $N_{lhd}$, set by the user. Besides $N_{lhd}$ points, the corners and the midpoints of each edges of the hypercube are also included in the initial design for a concrete representation of the boundaries of the search space.  The second step is to perform simulations to evaluate the extinction efficiencies of the initial sample points. Once the extinction efficiencies of the initial sample points are predicted, a surrogate model based on GP regression can be built using the initial dataset. Then, the accuracy of the GP regression model must be checked to understand whether it can sufficiently represent the underlying function or not.  For that, five-fold cross validation adopted by Ajdari and Mahlooji is used.  The five-fold validation relies on randomly dividing the sampled points into five subgroups that consist of approximately same number of points. One of these subgroups is temporarily left outside as a cross validation set, while a GP regression is performed using the sample points of each of the remaining 4 subgroups. The prediction error for each GP regression based on subgroup, $\mathit{j}$, can be defined in terms of a root relative square error (RRSE) using the cross validation set that is defined as,
		\begin{equation}
		RRSE_j = \sqrt{\frac{\sum_{i=1}^{N_{cv,j}} (y_{i,j} - \hat{y}_{i,j})^2}{\sum_{i=1}^{N_{cv,j}} (y_{i,j} - \bar{y}_{j})^2}}
		\label{eq:RRSE}
		\end{equation}
		where $y_{i,j}$ and $\hat{y}_{i,j}$ are the observation and GP prediction of the $i^{th}$ input point of $j^{th}$ subgroup. $N_{cv,j}$ is the number of elements that the subgroup has and
		\begin{equation}
		\bar{y}_j=\frac{1}{N_{cv,j}}\sum_{i=1}^{N_{cv,j}}{y_{i,j}}
		\label{eq:y_bar}
		\end{equation}

		The procedure described above is repeated for each of the five subgroups, and the total cross validation error, $E$, of the original set is computed as the average of five calculated RRSE as 
		
		\begin{equation}
		E=\frac{\sum_{j=1}^{5}{N_{cv,j} RRSE_j}}{\sum_{j=1}^{5}{N_{cv,j}}}
		\label{eq:Erav}
		\end{equation}		
		
		The original algorithm terminates when the total error diminishes beyond a prescribed tolerance, $\epsilon$, or the maximum sampling is achieved. However, a maximum sampling limitation is not utilized in our implementation and the algorithm terminates when the first criteria is satisfied.
		
		\begin{algorithm}
			\caption{Adaptive sequential design algorithm from \cite{Ajdari2014}}
			\begin{algorithmic} 
				\STATE \textbf{input}: Initial design points
				\STATE Perform the simulations of initial design
				\STATE Perform GP regression
				\STATE Check the stopping criteria
				\IF {stopping criteria is met}
				\STATE \quad Stop
				\ELSE
				\STATE \quad Add new point and perform its simulation
				\STATE \quad return to step where GP regression is performed
				\ENDIF
				\STATE \textbf{output}: GP model
			\end{algorithmic}
			\label{algorithm:adaptive_sequential_design_algorithm}
		\end{algorithm}

		The quality of an efficient adaptive sequential algorithm relies on how new training points are identified. The first step is to decompose the entire input domain into smaller sub-domains. The algorithm implemented uses Delaunay triangulation (DT) method that generates mesh of triangles using the sampled points as the vertices of the triangles. Then, the interior regions of the triangles generated by this strategy identify the unsampled regions inside the search domain. After unsampled regions are identified, the next step is to find the most promising region based on exploration and exploitation scores and to add a new point there. As mentioned before, the exploration strategy is responsible for good search space coverage, whereas the exploitation strategy identifies interesting regions of the output domain. Since each Delaunay triangle represents an unsampled region, the exploration score can be defined over their areas for a 2-D system, meaning that larger areas correspond to bigger exploration score, that is defined as,
		\begin{equation}
		\phi_i = \frac{A_i}{A_{max}} \times 10
		\label{eq:score_exploration}
		\end{equation}
		where $\phi_i$ and $A_i$ are the exploration score and the area of the $i^{th}$ triangle, respectively. The maximum area $A_{max}$ is calculated as $A_{max} = \max(A_i)$. For an $\mathit{n}$-dimensional system, $\mathit{n}$-dimensional volume can be used to calculate the exploration score.
		
		Exploitation score is a measure of variations among the extinction efficiencies of sample points. Then, the exploitation score for each triangle can be calculated as the average of the differences between extinction efficiency values of pair of sample points constituting the vertices. For a triangle with vertices referred as $A$, $B$ and $C$, the exploitation score, $\Delta_i$, is defined as, 
		\begin{align}
		&\delta_{i,1} = |y_{A,i} - y_{B,i}|	\\
		&\delta_{i,2} = |y_{A,i} - y_{C,i}|  \\
		&\delta_{i,3} = |y_{B,i} - y_{C,i}|  \\
		&\mathbf{\delta_i} = \sum_{j=1}^{3}{\delta_{i,j}}\\
		&\Delta_i = \frac{\mathbf{\delta_i}}{\mathbf{\delta_{max}}} \times 10
		\label{eq:score_exploitation}
		\end{align}
		
		Finally, the total score of a triangle can be calculated as a weighted sum of its exploration and exploitation scores.
		\begin{equation}
		\Theta_i = \alpha \phi_i + (1-\alpha) \Delta_i
		\label{eq:score_toal}
		\end{equation}
		where $\alpha$ is the weight parameter with $\alpha \in \left[0 \, 1\right]$. In the original algorithm of Ajdari and Mahlooji, $\alpha$ is also implemented as adaptively and it is changing in every iteration to take care of complex output surfaces \cite{Ajdari2014}. 
		
		When the total score of each triangle is computed, the most promising region is considered as the triangle with the highest score, and the next point that will be added is selected as the centroid of that triangle. Here, the algorithm of Ajdari and Mahlooji is modified slightly. Their algorithm is designed for continuous input variables. However, the total number of nanoparticle of an aggregate, $N_p$, can only have integer values in the problem considered. Therefore, once the final design points are determined the $N_p$ values of the final design points are rounded to the nearest integer. 
		
		In summary, this algorithm sequentially adds points to search space and a GP regression is fitted at every iteration. Then, the algorithm calculates the prediction error and compares it with prescribed tolerance level. Finally, the algorithm stops when $E<\epsilon$, and returns the GP regression model satisfying the predetermined tolerance criteria.

\section{Results and Discussion}

\subsection{\label{sec:choice_of_covariance_function}Choice of Covariance Function}
		According to Eqs. \ref{eq:gp_definition} and \ref{eq:gp}, GP is completely defined if the covariance function is specified satisfying a  zero mean function. Therefore, choosing an appropriate covariance function is crucial. Since the trends of the optical properties can easily be obtained using RDG-FA, the surface topology of $Q_{ext}$ can be analyzed before choosing a covariance function.
		
		Figure \ref{fig:fig1} shows that $Q_{ext}$ monotonically increases as $N_p$, and $d_p$  $\sigma$ increases, and there are no local extrema or discontinuity. The surfaces are very smooth and $Q_{ext}$ values of similar inputs do not  significantly vary. Therefore, a covariance function that correlates neighboring points could be a good choice.
		
		A suitable candidate might be the squared exponential covariance function with ARD defined in Eq. \ref{eq:gp_covSEard}. Another suitable option might be the Matern covariance function with ARD that is defined as,
		\begin{equation}
		k(\mathbf{x}_p, \mathbf{x}_q) = \sigma_f^2 \, f_d(r_d) \, \exp(-r_d)
		\label{eq:gp_covMaternard}
		\end{equation}
		where $d$ is a non-negative parameter with $f_1(t)=1, f_3(t)=1+t, f_5(t)=f_3(t)+t^2/3$, respectively. $r_d$ is the distance with  $r_d = \sqrt{ d (\mathbf{x}_p- \mathbf{x}_q)^T \mathbf{\Lambda}^{-2}  (\mathbf{x}_p- \mathbf{x}_q)}$ where $\mathbf{\Lambda}$ is a diagonal matrix with ARD parameters $\lambda_1,\lambda_2 ... \lambda_D$.
		
		Table \ref{tab:covariance_functions} shows the covariance functions used to build the surrogate models. The estimations based on these two covariance functions are analyzed and compared.	
		\begin{table}[htbp]
			\centering
			\caption{\label{tab:covariance_functions}The covariance functions studied in this work.}
			\begin{tabular}{ccccc}
				\hline
				Covariance & Description\\
				\hline
				1 & Squared exponential with ARD in Eq. \ref{eq:gp_covSEard}  \\
				2 & Matern with ARD in Eq. \ref{eq:gp_covMaternard} with $d=5$ \\
				\hline
			\end{tabular}
		\end{table}

	\subsection{\label{sec:effect_of_exploration_and_exploitation}Effect of Exploration and Exploitation}
		The effect of exploration and exploitation strategies can be better understood with a monodisperse example that has 2 variables, $\mathbf{x} = [N_p \, d_p \, \sigma=1.0]$. As mentioned earlier, exploration strategy aims to cover input domain as uniformly as possible whereas exploitation strategy aims to identify the most interesting regions such as local extremas or discontinuities. The weight parameter $\alpha$ in Eq. \ref{eq:score_toal} is used to adjust the contributions of these two strategies in determination of the total score of the Delaunay triangles. Since there is no local extrema or discontinuity in Fig. \ref{fig:fig1}, it can be concluded that the exploration strategy is the dominant factor in this study. 
		
		In order to see the effect of $\alpha$, the proposed method is executed with different $\alpha$ values starting with the same initial points. Then, the total number of final design points for a given tolerance $\epsilon$ can be used as a metric. The initial design points are sampled using LHD of size $N_{lhd}=10$. The corners and middle points of the boundaries of the input space  are added to 10 points, resulting 18 input points (10 from LHD, 4 corners and 4 middles). $Q_{ext}$ of every point is calculated using the RDG-FA. The proposed method is repeated 10 times for a given $\alpha$ and $\epsilon$ with 50 test points that are selected randomly from a uniform distribution with the parameter constraints considered in this study and used as a cross validation set. Finally, the procedure above is repeated for three different tolerance values ($\epsilon=10^{-3}, 5 \times 10^{-4}, 10^{-4}$) for two covariance functions (squared exponential and Matern).
		
\begin{figure}[h]
			\centering
			\fbox{\includegraphics[width=\linewidth]{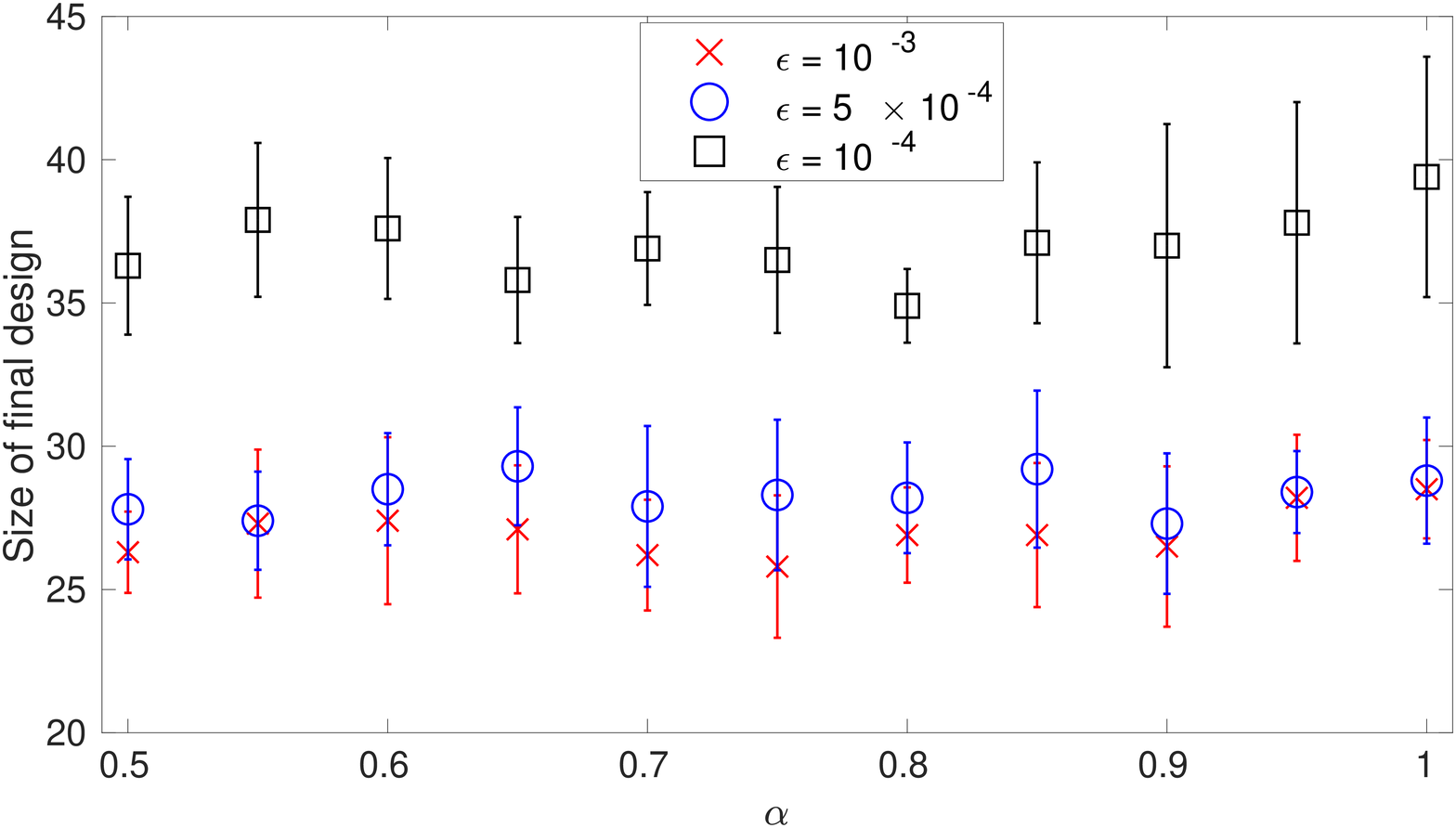}}
			\caption{Mean $\pm$ standard deviations of total number of sampled points for different tolerance levels for squared exponential covariance function.}
			\label{fig:fig2}
		\end{figure}

		Figure \ref{fig:fig2} shows the mean and standard deviations of the total number of points sampled by the proposed method for three different tolerance levels for squared exponential function with ARD distance measure. $\alpha$ values are started from 0.5 (meaning half exploration, half exploitation) with 0.05 increments up to 1 (meaning full exploration). It can be clearly observed that the total number of final design points remain approximately the same for each tolerance level. Expectedly, the total number of final design points increases as tolerance level decreases. The results for Matern covariance functions are also similar; hence, they are not presented here. Following these studies, $\alpha=0.7$ is chosen for the future analyses in this study. 
	
\begin{figure}[h]
			\centering
			\fbox{\includegraphics[width=\linewidth]{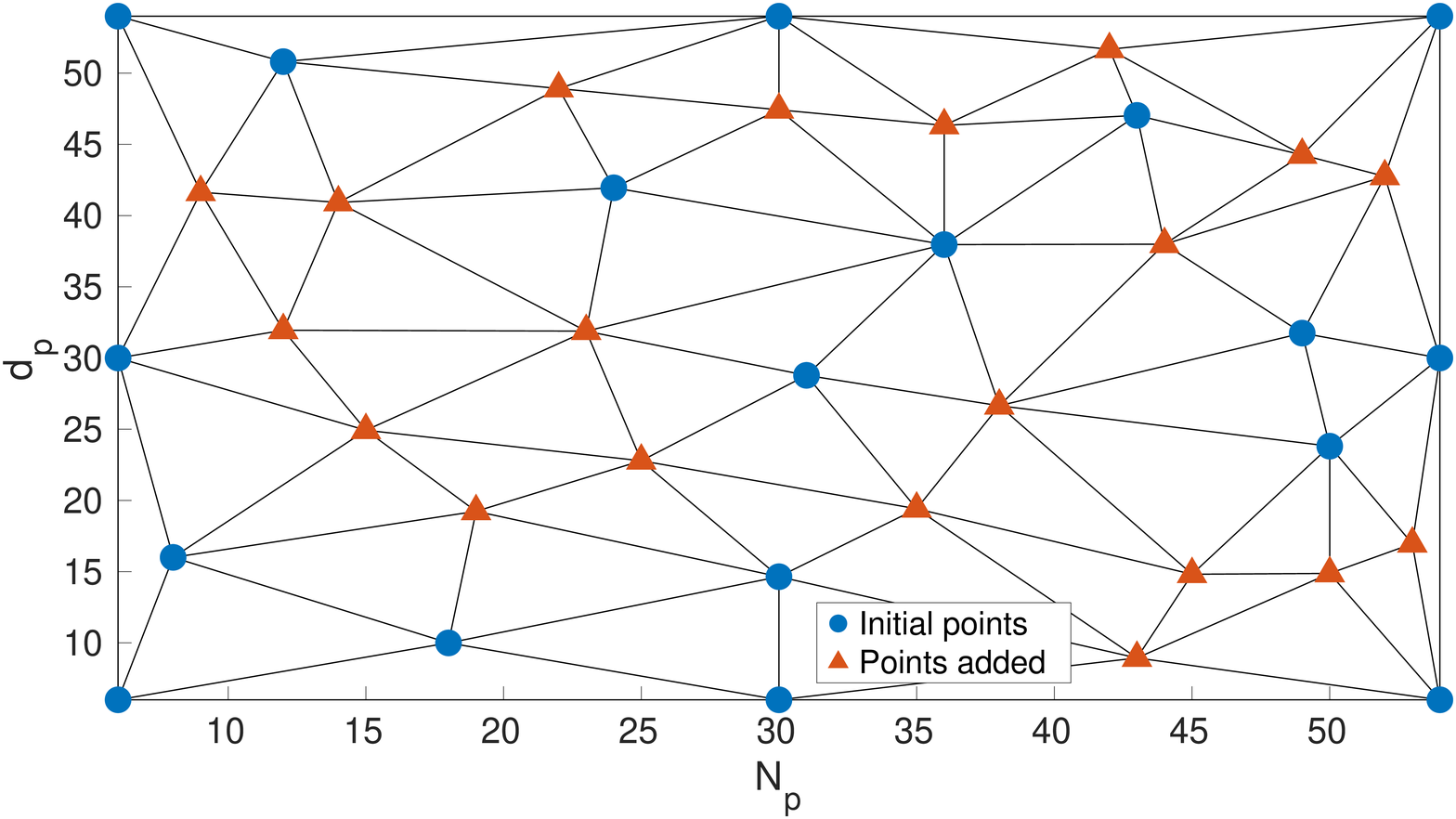}}
			\caption{Initial and final design points for $\alpha=0.7$ and $\epsilon=10^{-4}$ with squared exponential covariance function.}
			\label{fig:fig3}
		\end{figure}

		Figure \ref{fig:fig3} shows the Delaunay triangulation of the final design points as well as the initial design points for a monodisperse example when $\alpha=0.7$ and $\epsilon=10^{-4}$. As before, only the results for squared exponential covariance function are presented here. The blue circles in Fig. \ref{fig:fig3} correspond to 18 initial design points whereas the red triangles correspond to points added by the adaptive sequential design algorithm. It can be observed that the algorithm successfully samples new points from the unsampled regions, and covers the entire input domain. The union of blue circles and red triangles forms the final design points from which a GP regression can be made.

	\subsection{\label{sec:number_of_realizations_study}Number of Realizations Study}	
		Once an accurate surrogate model is built for RDG-FA, the final design points of this surrogate model will be used to built a surrogate model for DDA. While calculating the outputs of the final design points with DDSCAT, the optical properties of multiple aggregate realizations generated with same exact parameters must be averaged. Therefore, the number of adequate aggregate realizations must be identified for a typical parameter set.
		
		For that, the average value of $Q_{ext}$ is plotted as a function of the number of realizations. When the average reaches an acceptable value, then it can be concluded that the corresponding number of realization can be used while averaging. The acceptable value for $Q_{ext}$ depends on the accuracy of the model used; therefore, it can be determined by comparing DDSCAT with Mie theory. It is observed that the absolute error percentage of $Q_{ext}$ calculated by Mie theory and DDSCAT for a single sphere is smaller than $1.25\%$, and it decreases as $d_p$ increases. Based on this observation and considering the variations in optical properties, it is considered that an acceptable average value for $Q_{ext}$ lies within the interval of $\pm 2\%$ of the total average value. The average of $Q_{ext}$ is defined as,
		\begin{equation}
		\bar{Q}_{ext} = \frac{1}{N_r} \sum_{i=1}^{N_r} Q_{ext}
		\label{eq:qext_average}
		\end{equation}
		where $N_r$ is the number of realizations.
		
\begin{figure}[h]
			\centering
			\fbox{\includegraphics[width=\linewidth]{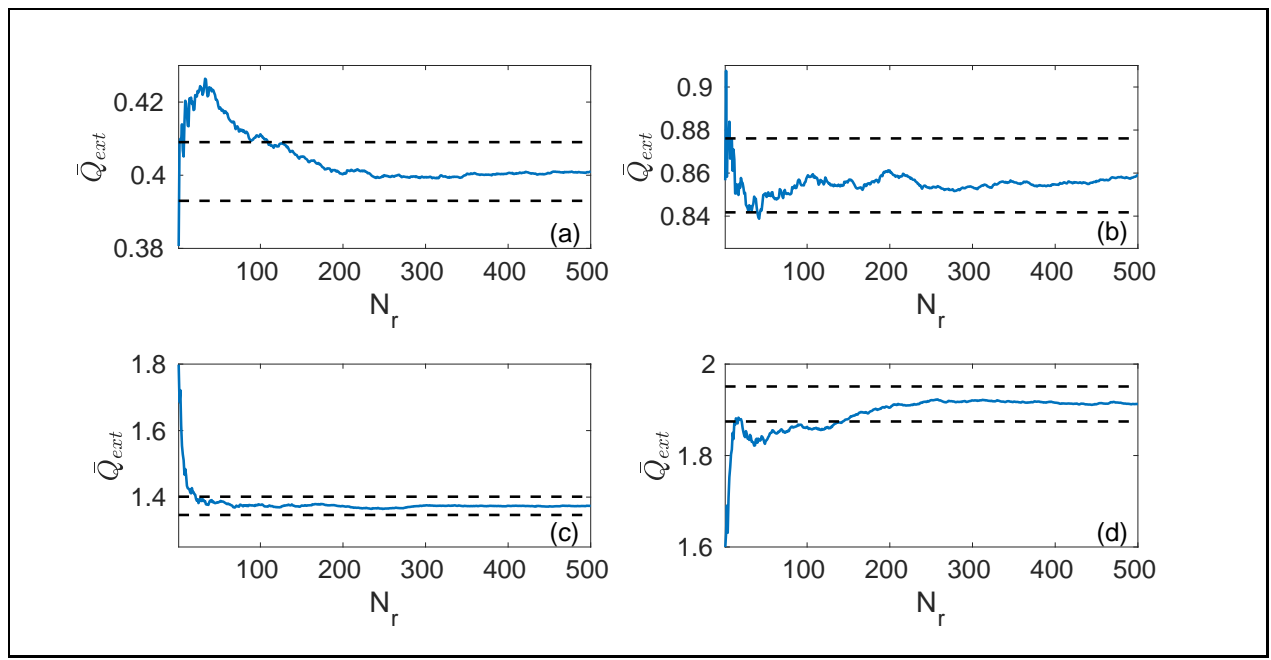}}
			\caption{$\bar{Q}_{ext}$ as a function of number of realizations of an aggregate with $N_p=30$, $\sigma=1.5$ for a) $d_p=12$ nm, b) $d_p=24$ nm, c) $d_p=36$ nm and d) $d_p=48$ nm.}
			\label{fig:fig4}
		\end{figure}

		Figure \ref{fig:fig4} shows $\bar{Q}_{ext}$ as a function of $N_r$ for four different $d_p$ values of an aggregate with $N_p=30$ and $\sigma=1.5$. The black dashed lines are  $\pm 2\%$ of $\bar{Q}_{ext}(N_r=500)$. It can be observed that $\bar{Q}_{ext}$ at $N_r=200$ lies within the prescribed limits for all $d_p$ values. The geometric standard deviation $\sigma=1.5$ corresponds to the widest lognormal distribution considered in this study. Similar plots are analyzed for smaller $\sigma$ values, and lower number of realizations are obtained from these analysis. Therefore, it can be safely concluded that averaging with $N_r=200$ realizations provides a good estimation for the optical properties.

	\subsection{\label{sec:gpr}GP Regression}		
		As explained before, the proposed method is first used for modeling $Q_{ext}$ with RDG-FA used to sample training points that will be used to build surrogate model for DDA. Therefore, three different surrogate models that correspond to three different tolerance levels $\epsilon$ are built with RDG-FA theory for each covariance function. The identified training points are then used to built a surrogate model for DDA. 
		
		Since it is hard to interpret the error definition based on root relative square error defined in Eq. \ref{eq:RRSE}, an additional percentage error is defined as,
		\begin{equation}
		E_i = \frac{y_i - \hat{y}_i}{y_i} \times 100
		\label{eq:tolerance_level}
		\end{equation}
		where $y_i$ and $\hat{y}_i$ are $Q_{ext}$ calculated from actual models (either RDG-FA or DDSCAT) and surrogate model, respectively.

	\subsubsection{RDG-FA}
		The proposed method is first used with RDG-FA to sample training points that will be used to build surrogate model for DDA. Therefore, 100 different surrogate models are built for three different tolerance levels $\epsilon=10^{-3}$, $\epsilon=5 \times 10^{-4}$ and $\epsilon=10^{-4}$ for each covariance function presented in Table \ref{tab:covariance_functions}. The accuracy of each model is tested with 100 randomly determined test points.
		
\begin{figure}[h]
			\centering
			\fbox{\includegraphics[width=\linewidth]{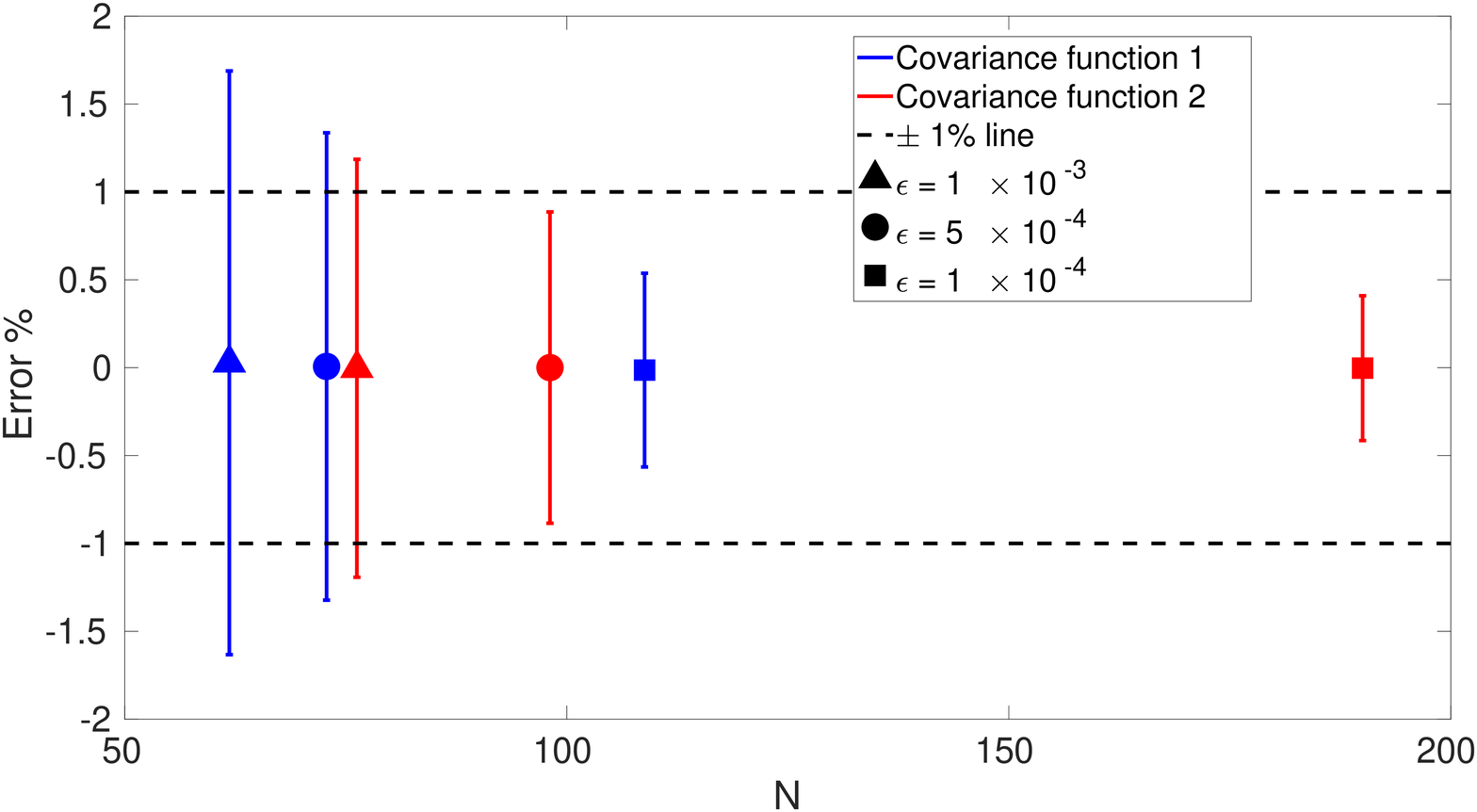}}
			\caption{The mean error with 99$\%$ confidence interval calculated with Eq. \ref{eq:tolerance_level} for different tolerance levels and covariance functions when $\alpha=0.7$. $N$ represents the average of total number of design points used to built the surrogate models.}
			\label{fig:fig5}
		\end{figure}

		Figure \ref{fig:fig5} shows the mean error of predicted $Q_{ext}$ with 99$\%$ confidence interval calculated with Eq. \ref{eq:tolerance_level} for three different tolerance levels $\epsilon=10^{-3}$, $5 \times 10^{-4}$ and $10^{-4}$ for both covariance functions considered in this study. The x-axis represents average number of points that each surrogate model has for the corresponding tolerance level and the covariance function. The black dashed lines are the $\pm 1\%$ error limits. It can be observed that the surrogate model built with Matern covariance function represents RDG-FA slightly better than the surrogate model built with squared exponential covariance function when $\epsilon=10^{-4}$. However, it is also observed that the average number of design points sampled with the Matern covariance function is larger than that of squared exponential covariance function for all tolerance levels. Therefore, both covariance functions are considered to built the surrogate models for DDA as they compete with each other in terms of accuracy and computational burden.

	\subsubsection{DDA}
		It is observed that the surrogate models, developed based on the covariance functions considered with a tolerance level of $\epsilon=10^{-4}$,  represent RDG-FA reasonably. Since RDG-FA captures the trend in $Q_{ext}$ correctly and there are no singularities in the functional values of $Q_{ext}$ as shown Fig. \ref{fig:fig1}, it can be presumed that the surrogate models built with these training points will also represent $Q_{ext}$ predicted DDA reasonably. Four different uniformly gridded databases (the increments of each parameter are given in Table \ref{tab:databases}) are also constructed to check whether the surrogate models are superior to conventional databases or not.

		\begin{table}[htbp]
			\centering
			\caption{\label{tab:databases}The databases considered in this work.}
			\begin{tabular}{ccccc}
				\hline
				Database & $\Delta N_p$ & $\Delta d_p$ [nm] & $\Delta \sigma_r$ & $N$ \\
				\hline
				1 	& 	12	&	12	&	0.1		& 150  \\
				2 	& 	12	&	12	&	0.05	& 275  \\
				3 	& 	6	&	6	&	0.1		& 486  \\
				4 	& 	6	&	6	&	0.05	& 891  \\
				\hline
			\end{tabular}
		\end{table}

		The accuracy of surrogate models and the databases constructed are tested using the outputs of 150 randomly selected test points. Among these test points, 50 are sampled outside the defined boundaries to check the extrapolation capabilities of both surrogate models and databases. These points are sampled with $d_p \in \left[  54 \,\, 66 \right]$,	$\sigma \in \left[  1.5 \,\,  1.6 \right]$ and	$N_p \in \left[  54 \,\, 66 \right]$. Cubic spline fit is used as the interpolation and extrapolation function for databases.
		
		The surrogate models for DDA are built by using the design points sampled based on RDG-FA with a tolerance of $\epsilon=10^{-4}$. Once the hyperparameters are identified using the training points the predictions of the surrogate models are compared with the predicted extinction efficiencies of the test points. The natural logarithm of the learned hyperparameters for each surrogate model are given in Table \ref{tab:log_hyperparameters}. 
		
		\begin{table}[htbp]
			\centering
			\caption{\label{tab:log_hyperparameters}The logarithm of the hyperparameters learned using the training points.}
			\begin{tabular}{ccccccc}
				\hline
				Covariance & $\ln \lambda_1$ & $\ln \lambda_2$ & $\ln \lambda_3$ & $\ln \sigma_f$ & $\ln \sigma_n$ & $N$\\
				\hline
				1	& 3.5113	& 4.2711	& -0.0016	& 0.6415	& -4.6976 	& 106	\\
				2	& 4.8669	& 5.5319	& 1.1091	& 1.4991	& -4.4229 	& 192	\\
				\hline
			\end{tabular}
		\end{table}		
		
\begin{figure*}
			\centering
			\fbox{\includegraphics[width=\linewidth]{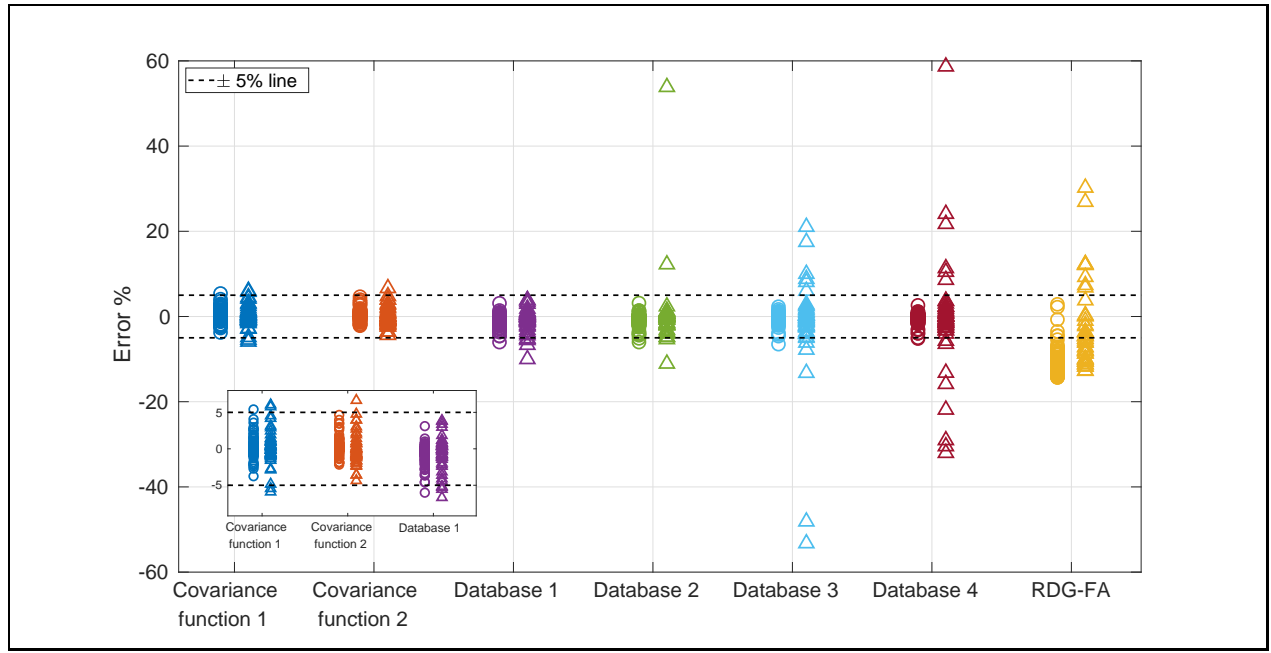}}
			\caption{Errors in the predictions calculated based on Eq. \ref{eq:tolerance_level} for two covariance functions, four databases and RDG-FA. Circles and triangles represent the errors for interpolated and extrapolated points, respectively. Inset is the zoom-in version of covariance functions and Database 1.}
			\label{fig:fig6}
		\end{figure*}

		Figure \ref{fig:fig6} shows a comparison of prediction errors for the points sampled within and outside the training domain. It can be observed that the errors in all approaches are similar to each other for the points sampled inside the training domain, meaning that all the considered approaches performs well while  interpolating.  For extrapolation, errors in the predictions of both covariance functions are similar to each other while the errors for covariance function 2 are less than those of covariance function 1. However, they both perform better than databases and the overall results indicate that the prediction based on covariance function 2 is slightly better than that of covariance function 1.  It can also be observed from Figure \ref{fig:fig6} that using databases result in particularly poor predictions when extrapolating. Moreover, the RDG-FA predictions are subject to significant error.

		It should be noted that the predictions obtained by the surrogate model for DDA are based on the training points sampled using RDG-FA.  The accuracy of the predictions might be increased if the training points are sampled using DDA instead of RDG-FA during the sampling stage.  However, it should also be noted that following that strategy would increase the computational cost.  Roughly, sampling 100 points with RDG-FA takes order of minutes whereas the time required to sample same number of points using DDA would be in the order of couple of weeks.

\subsubsection{Modeling $S_{11}$}
Aside from extinction efficiency, modeling directional properties of a scatterer that is strongly affected by its shape and size is also desired for many applications. In this section, the proposed method is used to build a preliminary surrogate model for one of the directional properties, $S_{11}$. The main objective here is to develop a surrogate model for the error of RDG-FA predictions of $S_{11}$ that are calculated based on predictions of DDSCAT so that this surrogate model can be used to correct the erroneous RDG-FA predictions. Here, we use the same training dataset that we used in previous sections with a minor modification.

In this section, the input points $\mathbf{x} \in \mathcal{R}^4$ with $\mathbf{x} = [ N_p \, d_p \, \sigma  \, \theta]$, and the outputs are the corresponding $S_{11}$ values. It must be noted that the preliminary surrogate models in this section are trained with the the previous datasets, which are formed using $Q_{ext}$.

There are some experimental constraints and limitations in measuring the angular properties. The measurements of scattered light at the small angles are highly disturbed as distinguishing the incident light from the scattered light is difficult without using well designed instrumentation \cite{Heinson2015}.  For large scattering angles, the incident light gets blocked due to finite size of the measurement device, as explained by \cite{Hovenier2003}. Besides, the increments under $16^{\circ}$ is enough to obtain reasonable results as reported by Ericok and Erturk \cite{Ericok2017}. Therefore, the measurements for $S_{11}$ is taken between $5^{\circ}$ to $165^{\circ}$ with an increment of $8^{\circ}$. The percentage error to be analyzed between the two methods is defined as

		\begin{equation}
		S_{11,error} = \frac{S_{11,DDSCAT} - S_{11,RDG-FA}}{S_{11,RDG-FA}} \times 100
		\label{eq:S11_error}
		\end{equation}

\begin{figure}[h]
	\centering
    \fbox{\includegraphics[width=\linewidth]{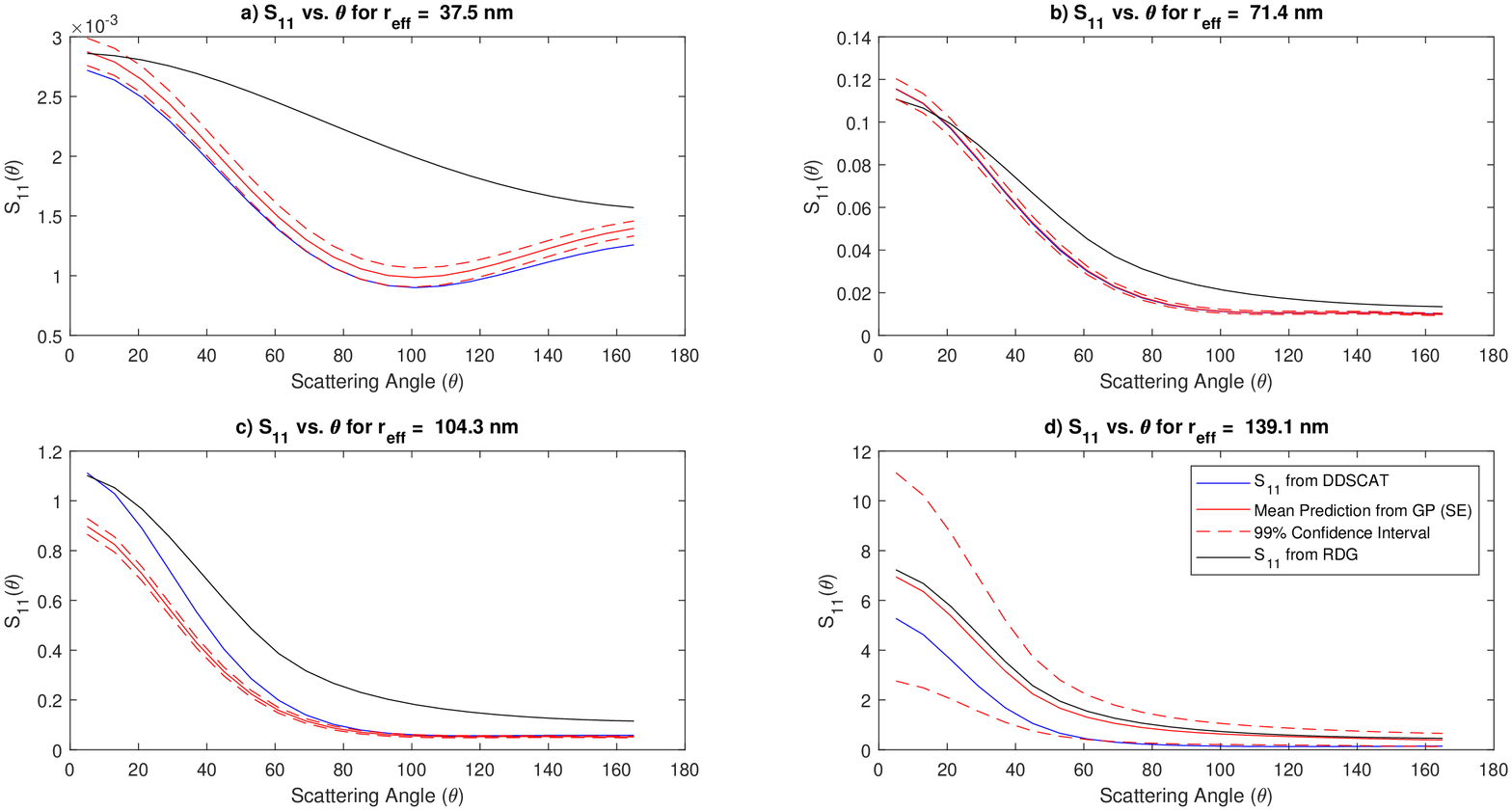}}
    \caption{First element of Mueller matrix ($S_{11}$) versus the scattering angle at various effective radius values for four test points a, b, c) Interpolated points and d) Extrapolated Point}
    \label{fig:fig7}
    \end{figure}

	Gaussian process is applied using Matern and squared exponential covariance function with ARD for modeling the error that is then used to correct the RDG-FA estimations. Figure \ref{fig:fig7} shows $S_{11}(\theta)$ calculated using DDSCAT, RDG-FA and corrected RDG-FA with the surrogate error model calculated based on the squared exponential covariance function for four cases with different effective aggregate sizes. The aggregate sizes are chosen so that effective radii are all greater than 35 nm, which is the lowest possible characterization limit of soot aggregates for $\lambda=532$ nm \cite{Ericok2018a}.  It is observed that the results with Matern and squared exponential covariance functions are very similar; therefore, only the predictions of squared exponential covariance function are presented. Figure \ref{fig:fig7} shows that correcting RDG-FA with a surrogate error model using squared exponential covariance function improves the RDG-FA predictions drastically, and the corrected predictions are in agreement with DDSCAT predictions.

	A prediction error is defined in order to quantify and compare the performance of the surrogate model corrected RDG-FA predictions using the two covariance functions with RDG-FA predictions and predictions based on a database using cubic spline interpolation.  The prediction error, $PE_i$, is defined based on the $S_{11}(\theta)$ vector obtained for a test point $\mathbf{x}=[N_p \, d_p \, \sigma]$ as
	
		\begin{equation}
		PE_i= \frac{\left\lVert S_{11,prediction} - S_{11,DDSCAT}\right\rVert}{\left\lVert S_{11,DDSCAT}\right\rVert} \times 100
		\label{eq:prediction_error}
		\end{equation}
		where $S_{11,prediction}$ is one of the aforementioned four methods and $S_{11,DDSCAT}$ is actual prediction of DDSCAT.
	
\begin{figure}[h]
	\centering
\fbox{\includegraphics[width=\linewidth]{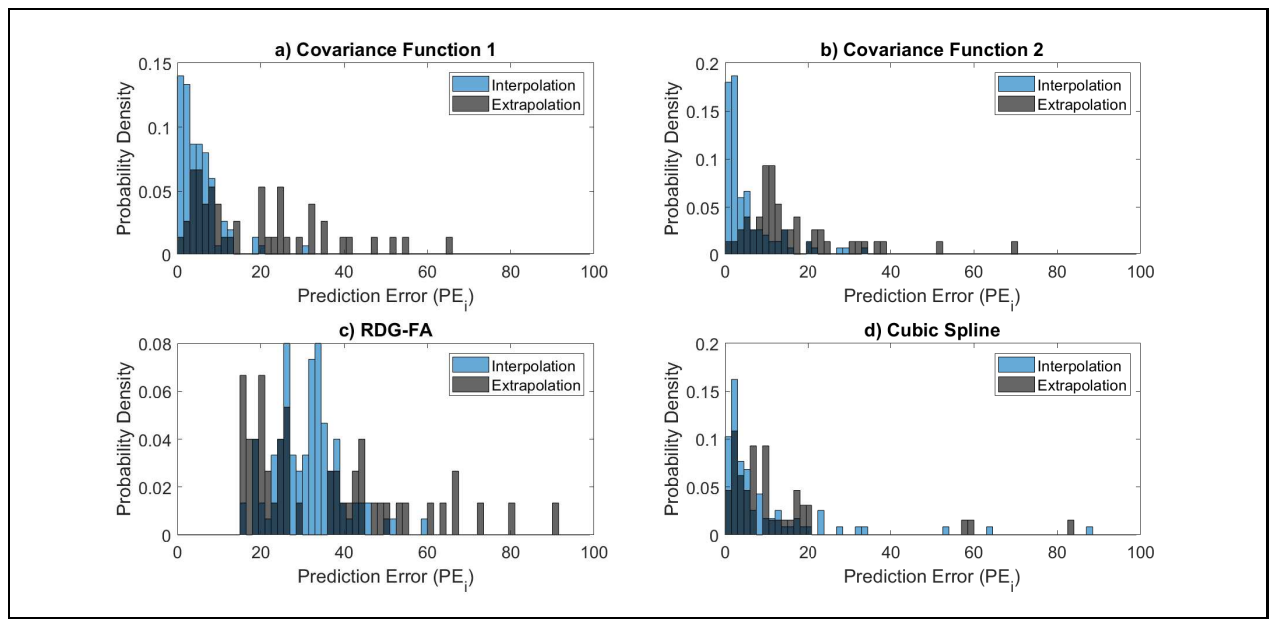}}
\caption{Comparison of prediction errors of the methods applied.}
\label{fig:fig8}
\end{figure}

	Figure \ref{fig:fig8} shows that corrected RDG-FA with the two covariance functions performs better for both interpolating and extrapolating test points compared to uncorrected RDG-FA predictions. It can be observed that the errors for both covariance functions and cubic spline interpolation through database are similar to each other for the interpolated test points, meaning that all three approaches performs well while interpolating. It can be observed that the prediction errors of covariance function 2 are slightly less than those of covariance function 1 in average. However, they both perform better than the cubic spline interpolation from  the Database 2. It should be noted that a few test points with prediction error values beyond $100\%$ are not presented in the Fig. \ref{fig:fig8}.d for cubic spline interpolation from the database. These outliers correspond to the fractals where effective radius is lower than the lowest characterization limit, which was reported as 35 nm for 532 nm \cite{Ericok2018a}. Below this characterization limit the order of magnitude of $S_{11}$ values is lower than the measurement uncertainty.

	From the comparison of Gaussian process predictions and predictions of DDSCAT for the test points considered, Gaussian process is found to be successful in capturing the values and general trend of the DDSCAT results.  While the confidence interval is relatively narrow in most of the cases, in some cases a wide confidence intervals are observed. Based on these preliminary results, it can be concluded that using Gaussian process based supervised machine learning technique to predict the directional properties of the aggregates is a promising method. However, the accuracy of the method requires further improvements that can be achieved by sampling the training points based on $S_{11}$ values, or improving the sampling algorithms.

\section{Conclusions}
In this study, a systematic procedure to construct a fast and accurate surrogate model for the optical properties calculated by DDA for an ensemble of nanoparticle aggregates are introduced. The aggregates are assumed to have same number of nanoparticles with a lognormal particle size distribution. The aggregates are generated using the Filippov's cluster-cluster algorithm. The optical property considered for an ensemble is the extinction efficiency, and it is calculated using DDSCAT by averaging the efficiencies of 216 orientations and 200 aggregate realizations.
	
	The surrogate models are built with GP regression, and the input points for the training dataset are sampled using the adaptive sequential design algorithm proposed by Ajdari and Mahlooji \cite{Ajdari2014}. The covariance functions considered in this work are the squared exponential and Matern covariance functions with ARD parameters. The strength and weaknesses of the proposed methodology are first tested with RDG-FA, and it is observed that accurate surrogate models can be built for RDG-FA. 
	
	The input points sampled with RDG-FA are also used to built surrogate models for DDA. First, the extinction efficiencies of the input points are calculated using DDSCAT. Then, the surrogate models are built using these datasets. Four different uniformly gridded databases are also built for comparison. It is observed that the surrogate models based on squared exponential and Matern covariance functions are superior to databases in terms of both accuracy and the total number of input points they require. Thus, a very time consuming process such as building an accurate database is reduced by a considerable amount. The surrogate model built with Matern covariance function produces accurate predictions than the surrogate model built with squared exponential covariance function. However, it is also observed that the former requires more training points than the latter, meaning that it is more computationally demanding.
	
	Finally, the directional properties of soot aggregates can also be modeled with this approach. A preliminary surrogate model for the error in $S_{11}$ is built to correct the predictions of RDG-FA method. Thus, the speed of RDG-FA method is combined with the accuracy of DDA. Although the preliminary results are promising, a more comprehensive study is required to capture all the details in estimating the directional properties.
	
	In this study, the number of the unknown input parameters are limited to three while there are uncertainties in regards to other parameters such as complex refractive index or fractal parameters. Moreover, effects of particle overlapping and necking is ignored.  However, the proposed method can easily be extended to consider the uncertainties in refractive index, fractal parameters, and overlapping and necking effects that effect the optical behavior considerably.  Although extinction efficiency and S11 are considered in this study, the proposed methodology is flexible enough to be applied to other properties.


\bibliographystyle{unsrt}

\bibliography{ericok_et_al_jqsrt_2019}


\end{document}